\documentclass[sigconf,balance]{acmart}
\usepackage{graphicx}
\usepackage{subfiles}

\usepackage{todonotes}
\usepackage{booktabs}
\usepackage{multirow}
\PassOptionsToPackage{hyphens}{url}\usepackage{hyperref}

\usepackage{dsfont}

\usepackage{amsmath}
\usepackage{braket}%

\DeclareMathOperator{\E}{\mathbb{E}}

\usepackage{caption}
\usepackage{subcaption}
\usepackage{csquotes}

\usepackage{xcolor}

\AtBeginDocument{%
  \providecommand\BibTeX{{%
    \normalfont B\kern-0.5em{\scshape i\kern-0.25em b}\kern-0.8em\TeX}}}

\copyrightyear{2024}
\acmYear{2024}
\setcopyright{rightsretained}
\acmConference[SIGIR '24]{Proceedings of the 47th International ACM SIGIR Conference on Research and Development in Information Retrieval}{July 14--18, 2024}{Washington, DC, USA}
\acmBooktitle{Proceedings of the 47th International ACM SIGIR Conference on Research and Development in Information Retrieval (SIGIR '24), July 14--18, 2024, Washington, DC, USA}
\acmDOI{10.1145/3626772.3657943}
\acmISBN{979-8-4007-0431-4/24/07}

\begin{document}

\title{Language Fairness in Multilingual Information Retrieval}

\author{Eugene Yang}
\affiliation{%
  \institution{Johns Hopkins University}
  \city{Baltimore}
  \state{Maryland}
  \country{USA}
}
\email{eugene.yang@jhu.edu}

\author{Thomas Jänich}
\affiliation{%
  \institution{University of Glasgow}
  \city{Glasgow}
  \country{United Kingdom}
}
\email{t.jaenich.1@research.gla.ac.uk}

\author{James Mayfield}
\affiliation{%
  \institution{Johns Hopkins University}
  \city{Baltimore}
  \state{Maryland}
  \country{USA}
}
\email{mayfield@jhu.edu}

\author{Dawn Lawrie}
\affiliation{%
  \institution{Johns Hopkins University}
  \city{Baltimore}
  \state{Maryland}
  \country{USA}
}
\email{lawrie@jhu.edu}

\begin{abstract}
Multilingual information retrieval (MLIR) considers the problem of ranking documents in several languages
for a query expressed in a language that may differ from any of those languages. 
Recent work has observed that approaches such as 
combining ranked lists representing a single document language each
or using multilingual pretrained language models
demonstrate a preference for one language over others.
This results in systematic unfair treatment of documents in different languages. 
This work proposes a language fairness metric
to evaluate whether documents across different languages are fairly ranked
through statistical equivalence testing using the Kruskal-Wallis test. 
In contrast to most prior work in group fairness,
we do not consider any language to be an unprotected group.
Thus our proposed measure, PEER (Probability of Equal Expected Rank), is the
first fairness metric specifically designed to capture the language
fairness of MLIR systems. 
We demonstrate the behavior of PEER on artificial ranked lists. 
We also evaluate real MLIR systems on two publicly available benchmarks
and show that the PEER scores align with prior analytical findings on MLIR fairness. 
Our implementation is compatible with \texttt{ir-measures}
and is available at \url{http://github.com/hltcoe/peer\_measure}.

\end{abstract}

\ccsdesc[500]{Information systems~Multilingual and cross-lingual retrieval}
\ccsdesc[500]{Information systems~Evaluation of retrieval results}

\keywords{language fairness, multilingual retrieval, statistical testing}

\maketitle

\section{Introduction}

Multilingual information retrieval searches a multilingual document collection
and creates a unified ranked list for a given query~\cite{clef2001, clef2002, clef2003, lawrie2023neural, mulm, huang2023soft}. 
In tasks like navigational search~\cite{broder2002taxonomy}, known item retrieval~\cite{lee2006known, arguello2021tip}, and retrieval for question-answering~\cite{chen2017reading, kwon2018controlling},
the user only needs a handful or just one relevant document to satisfy the information need,
and the language of that document does not matter.
In contrast, users interested in gaining a broad problem understanding
prefer seeing how coverage varies across languages.

Analysis of retrieval results has shown that MLIR systems
often show a preference for certain languages~\cite{lawrie2023neural, huang2023soft};
we call this the \textit{MLIR Fairness problem.}
Such preference can bias a user's understanding of the topic~\cite{white2013beliefs, schulz2000biased}. 
This problem is particularly apparent in models built on top of multilingual pretrained language models (mPLM)~\cite{lawrie2023neural},
which inherit bias from the text used to build them~\cite{choudhury2021linguistically, kassner2021multilingual}.
This paper presents a new metric to allow quantitative study of MLIR Fairness.

Prior work in fairness evaluation focuses on either individual or group fairness~\cite{zehlike2021fairness}.
Individual fairness ensures that similar documents receive similar treatment;
this often corresponds to a Lipschitz condition~\cite{biega2018equity,dwork2012fairness}.
Group fairness ensures that a protected group receives treatment at least as favorable as unprotected groups~\cite{zehlike2017fa,zehlike2022fair,sapiezynski2019quantifying}.
Group fairness metrics designed for protecting specific groups are not directly applicable to the MLIR fairness problem
because the latter has no protected language;
we want all languages to be treated equally in a ranked list.

To operationalize our notion of MLIR fairness, we propose the Probability of Equal Expected Rank (PEER) metric. 
By adopting the Kruskal-Wallis $H$ test,
which is a rank-based, non-parametric variance analysis for multiple groups,
we measure the probability that documents of a given relevance level for a query are expected to rank at the same position
irrespective of language.
We compare PEER to previous fairness metrics,
and show its effectiveness on synthetic patterned data,
on synthetic assignment of language to real retrieval ranked lists,
and on system output for the CLEF 2003 and NeuCLIR 2022 MLIR benchmarks.

\section{Related Work}

There is no universally accepted definition of fairness.
This paper views languages as groups within a ranking,
and characterizes MLIR Fairness as a group fairness problem.

Existing group fairness metrics fall into two categories:
those that assess fairness independent of relevance,
and those that take relevance into account.
\textit{Ranked group fairness}, based on statistical parity proposed by \citet{zehlike2017fa,zehlike2022fair},
demands equitable representation of protected groups in ranking
without explicitly considering relevance through statistical testing.

Attention Weighted Ranked Fairness (AWRF),
introduced by \citet{sapiezynski2019quantifying}, %
compares group exposure at certain rank cutoffs
against a pre-defined target distribution.
It uses the same distribution for both relevant and nonrelevant documents.
This means for example that if utility is defined as finding the most relevant documents,
a system can gain utility by including more documents from the language with the most relevant documents
early in the rankings.
In doing so, more nonrelevant documents from that language are placed above relevant documents from the other languages.
From a fairness perspective, this should be penalized as unfair.
Our proposed metric does not rely on a target distribution,
so it does not suffer from this utility/fairness tradeoff.

Among metrics that incorporate relevance,
\citet{singh2018fairness} introduced the Disparate Treatment Ratio,
which measures the equality of exposure of two groups.
This metric is not well suited to MLIR though,
since it handles only two groups.
Adjacent to fairness,
\citet{Clarke2008NoveltyAD} extended Normalized Discounted Cumulative Gain (nDCG)
to incorporate diversity. 
Their metric, $\alpha$-nDCG, assigns document weights based on both relevance and diversity. %
Diversity though applies to \textit{user} utility where fairness applies to \textit{documents}
(in our case, the languages of the returned documents)~\cite{castillo2019fairness}.
We nonetheless report $\alpha$-nDCG to contextualize our results.

Related work on fairness over sequences of rankings~\cite{morik2020controlling, diaz2020evaluating}
requires both more evidence and distributional assumptions
compared to fairness of a specific ranking.
While similar, our method assumes 
the position of each document is a random variable.

\section{Probability of Equal Expected Rank}

In this section, we describe the proposed measure -- PEER: Probability of Equal Expected Rank. 
We first introduce our notation and the fairness principle,
followed by forming the statistical hypothesis of the system's fairness across document languages. 
Finally, we define PEER as the $p$-value of the statistical test. 

Let $d_i\in\mathcal{D}$ be the $i$-th document in the collection $\mathcal{D}$ of size $N$. 
We define the language that $d_i$ is written in as $l_{d_i}\in\{\mathcal{L}_1, ... \mathcal{L}_M\}$. 
For convenience, we define the set $L_j = \Set{ d_i  |  l_{d_i} = \mathcal{L}_j }$ to be all documents in language $\mathcal{L}_j$.

For a given query $q\in \mathcal{Q}$, we define the degree of document $d_i$ being relevant (or the relevance grade) to the query $q$ as $y^q_i\in\{\mathcal{R}^{(0)}, \mathcal{R}^{(1)}, ..., \mathcal{R}^{(K)}\}$,
where $R^{(0)}$ indicates not relevant and $R^{(K)}$ is the most relevant level,
i.e., graded-relevance with $K$ levels. 
Similarly, we define the set $R^{(q,k)} = \Set{ d_i | y^q_i = \mathcal{R}^{(k)} }$ 
to be all documents at the $\mathcal{R}^{(k)}$ relevance level.
Furthermore, we define the documents in $\mathcal{L}_j$ with relevance level $\mathcal{R}^{(k)}$ for a query $q$ as 
$D_j^{(q, k)} = L_j \cap R^{(q,k)}$. 

In this work, we consider a ranking function $\pi: \mathcal{D}\times\mathcal{Q} \rightarrow [1...N]$ that produces the rank $r^q_i\in[1...N]$.

\subsection{Fairness through Hypothesis Testing}

We define MLIR fairness using the following principle:
\textit{Documents in different languages with the same relevance level, in expectation,
should be presented at the same rank. }
We measure the satisfaction level of this principle by treating it as a testable hypothesis. 

For relevance level $\mathcal{R}^{(k)}$, assuming $r^q_i$ is a random variable over $[1...N]$,
we implement the principle
using the null hypothesis: 
\begin{equation}
    H_0: \E_{d_i\in D_a^{(q, k)} }[r^q_i] = \E_{d_j\in D_b^{(q, k)}} [r^q_j]
    \,\,\,\, \forall \mathcal{L}_a \neq \mathcal{L}_{b}, 
\end{equation}
which is the equivalence of the expected rank among documents in each language with the given relevance level and given query $q$. 

Such null hypotheses can be tested with the Kruskal-Wallis $H$ test (K-W test)~\cite{kruskal1952use}. 
The null hypothesis of this test is that all groups have the same mean
(i.e., equivalent mean ranks). %
The K-W test is like a non-parametric version of the ANOVA F-test,
which tests whether each group (languages in our case) comes from the same distribution. 
Since the K-W test does not assume any particular underlying distribution,
it uses the ranking of the data points to make this determination.
Unlike prior work such as \citet{zehlike2022fair} that assumes a binomial distribution for each document over the groups,
not assuming the distribution of the query-document scores used for ranking and instead operating directly on ranks
yields a robust statistical test. 

Conceptually, the test statistics $H$ for the K-W test
is the ratio between the sum of group %
rank variance
and the total rank variance.
The variance ratio obeys a chi-squared distribution;
we use its survival function to derive the $p$-value. 
Specifically, we can express the test statistic $H$ as 
\begin{align}
    &H = \left(|R^{(q,k)}|-1\right) \frac{
        \sum_{j=1}^M \left| D_j^{(q, k)}\right| \left(\Bar{r}^{q,k}_j - \Bar{r}\right)^2
    }{
        \sum_{j=1}^M \sum_{d_i \in D_j^{(q, k)} } \left( r^q_i - \Bar{r} \right)^2
    } \\
    &\text{where } \Bar{r}^{q,k}_j  = \frac{1}{|D_j^{(q, k)}|} \sum_{d_i\in D_j^{(q, k)} } r^q_i \\
    &\text{and } \Bar{r} = \frac{1}{|R^{(q,k)}|} \sum_{j=1}^M \sum_{d_i \in D_j^{(q, k)} } r^q_i
\end{align}
for a given query $q$ and relevance level $\mathcal{R}^{(k)}$.
Recall that $R^{(q,k)}$ and $D_j^{(q, k)}$ are sets.
For each query $q$ and given relevance level, we report the $p$-value of the K-W test,
which is the \underline{P}robability of documents in all languages with given relevance level having \underline{E}qual \underline{E}xpected \underline{R}ank,
by comparing the $H$ statistic against a chi-squared distribution with $M-1$ degrees of freedom.
The $p$-value provides us with the probability that documents in different languages are ranked fairly
within a given relevance level. 
We denote the $p$-value for a given query $q$ and a relevance level $\mathcal{R}^{(k)}$ as $p^{(q, k)}$.

Our fairness notion is similar to the one proposed by \citet{diaz2020evaluating}.
However, we operationalize the principle by treating each document as a sample from a distribution given the language and relevance level,
instead of assuming the entire ranked list is a sample from all possible document permutations. 

\begin{figure*}
    \centering
    \includegraphics[width=\textwidth]{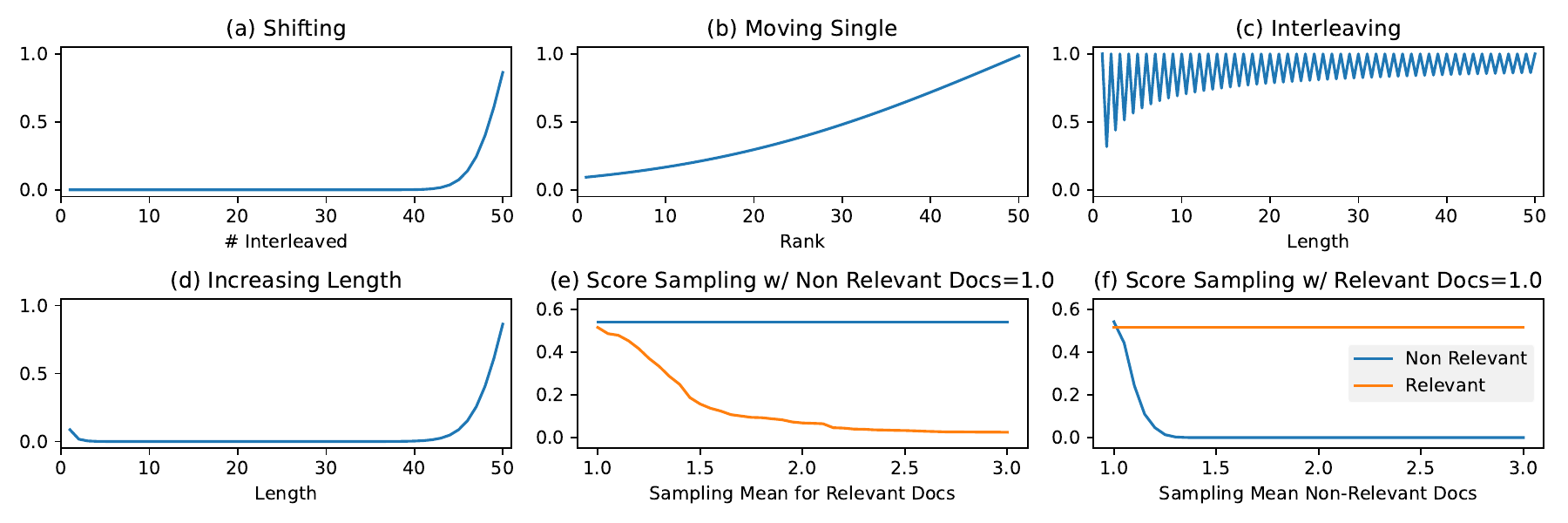}
    \vspace{-2em}
    \caption{Ranked lists with different fairness patterns between two languages and binary relevance.}\label{fig:diff-pattern}
\end{figure*}

\subsection{Fairness at Each Relevance Level}

The impact of unfairly ranking documents in different languages may differ at each relevance level.  
Such differences can be linked to a specific user model or application. %
For example, for an analyst actively seeking information for which each language provides different aspects,
ranking nonrelevant documents of a particular language at the top does not degrade fairness;
finding disproportionately fewer relevant documents in a certain language, on the other hand,
may yield biased analytical conclusions.

In contrast, for a user seeking answers to a specific question
who views the language as just the content carrier,
reading more nonrelevant documents from a language may degrade that language's credibility,
leading the user eventually to ignore all content in that language. 
In this case, we do consider the language fairness of the ranking of nonrelevant documents;
in the former case we do not.

To accommodate different user models, we define the PEER score as a linear combination of the $p$-value of each relevance level $\mathcal{R}^{(k)}$. 
Let $w^{(k)}\in [0, 1]$ be the weights and $\sum_{k=1}^K w^{(k)} = 1$, 
the overall weighted \textbf{PEER} for query $q$ is
\begin{equation}
    PEER^{(q)} = \sum_{k=1}^K w^{(k)} p^{(q, k)}
\end{equation}

\subsection{Rank Cutoff and Aggregation}

While a ranking function $\pi$ ranks each document in collection $\mathcal{D}$,
in practice, a user only examines results up to a certain cutoff. 
Some IR effectiveness measurements consider elaborate browsing models,
such as exponential-decreasing attention in Ranked-biased precision (RBP)~\cite{moffat2008rank}
or patience-based attention in expected reciprocal rank (ERR)~\cite{chapelle2009expected};
user behavior though is perpendicular to language fairness,
so we consider only a simple cutoff model. 

With a rank cutoff $X$,
we treat only the top-$X$ documents as the sampling universe for the K-W test.
However, since disproportionately omitting documents of a certain relevance level is still considered unfair,
before conducting the hypothesis test, 
we concatenate unretrieved documents 
(or those ranked below the cutoff)
at that relevance level
to the ranked list,
assigning them a tied rank of $X+1$. 
This is optimistic, since these documents might rank lower in the actual ranked list.
However, this provides a robust penalty for any ranking model that provides only a truncated ranked list. 
We define the $p$-value as 1.0
when no document is retrieved at a given relevance level in spite of their presence in the collection;
from the user perspective, no document at that level is presented,
so it is \textit{fair} (albeit ineffective) across languages.
We denote the weighted $p$-value calculated on the top-$X$ documents as $PEER^{(q)}@X$.

Overall, we report the average weighted PEER over all queries at rank $X$,
i.e., $PEER@X = |\mathcal{Q}|^{-1} \sum_{q\in\mathcal{Q}} PEER^{(q)}@X$.
Since we treat each document as a random variable of position in a ranked list,
what we are measuring is \textit{how likely} a system is fair between languages
instead of how fair each ranked list is. 
A higher PEER score indicates that the measured MLIR system is \textit{more likely}
to place documents written in different languages but with the same relevance level at similar ranks.

\section{Experiments and Results}

\begin{table*}
\renewcommand{\b}[1]{\textbf{#1}}
\newcommand{\andcg}{$\alpha$-nDCG}
\caption{Effectiveness and fairness results. Both AWRF and PEER exclude nonrelevant documents. They are removed in AWRF calculation. For PEER, the importance weights of nonrelevant documents are set to 0.  }\label{tab:main}
\centering

\begin{tabular}{l|l|c|ccc|c|ccc}
\toprule
\multirow{2.5}{*}{Collection}
&   Rank Cutoff &           \multicolumn{4}{c|}{20}     &     \multicolumn{4}{c}{1000} \\
\cmidrule{2-10}
&       Measure &    nDCG &  \andcg &    AWRF &    PEER &  Recall &  \andcg &    AWRF &    PEER \\
\midrule
\multirow{5}{*}{CLEF 2003} 
& QT >> BM25    &   0.473 &   0.444 &   0.513 &   0.239 &   0.743 &   0.579 &   0.788 &   0.202 \\
& DT >> BM25    &   0.636 &   0.640 &   0.623 &   0.243 &   0.857 &   0.747 &   0.895 &   0.299 \\
& DT >> ColBERT &\b{0.669}&\b{0.674}&\b{0.658}&   0.293 &\b{0.889}&\b{0.768}&\b{0.904}&   0.328 \\
& ColBERT-X ET  &   0.591 &   0.592 &   0.610 &   0.215 &   0.802 &   0.695 &   0.845 &   0.327 \\
& ColBERT-X MTT &   0.643 &   0.658 &   0.649 &\b{0.318}&   0.827 &   0.748 &   0.860 &\b{0.362}\\
\midrule
\multirow{5}{*}{NeuCLIR 2022} 
& QT >> BM25    &   0.305 &   0.447 &   0.537 &   0.453 &   0.557 &   0.569 &   0.752 &   0.383 \\
& DT >> BM25    &   0.338 &   0.448 &   0.542 &\b{0.497}&   0.633 &   0.580 &   0.809 &   0.421 \\
& DT >> ColBERT &\b{0.403}&   0.539 &\b{0.635}&   0.449 &\b{0.708}&\b{0.652}&\b{0.842}&\b{0.426}\\
& ColBERT-X ET  &   0.299 &   0.447 &   0.578 &   0.458 &   0.487 &   0.561 &   0.745 &   0.421 \\
& ColBERT-X MTT &   0.375 &\b{0.545}&   0.621 &   0.425 &   0.612 &   0.644 &   0.786 &   0.386 \\

\bottomrule
\end{tabular}

\end{table*}

\subsection{Synthetic Data}

To demonstrate PEER behavior,
we create ranked lists of two languages and binary relevance with four methods,
each creating lists from very unfair to very fair.
Results are illustrated in Figure~\ref{fig:diff-pattern}.
\textit{Shifting} starts with all documents in one language ranking higher than those in the other,
and slowly interleaves them until the two are alternating.
Figure~\ref{fig:diff-pattern}(a) shows for fifty documents that
when no documents are interleaved (left), fairness is low, with PEER close to 0.
As alternation increases, the PEER score increases.
In \textit{Moving Single,} the ranked list consists entirely of one language except for one document.
That single document moves from the top (unfair) to the middle of the ranking (fair).
In Figure~\ref{fig:diff-pattern}(b) with 99 majority language documents,
the PEER scores increase as the singleton moves from the top to the middle.
Figure~\ref{fig:diff-pattern}(c) shows \textit{Interleaving,}
in which the languages alternate and the number of retrieved documents slowly increases.
With odd lengths the highest and the lowest ranked documents are in the same language,
giving them the same average and 1.0 PEER scores.
With even lengths, one language has a slightly higher rank than the other.
The difference shrinks with longer ranked lists, resulting in increased PEER scores. 
In \textit{Increasing Length}, 100 retrieved documents comprise first an alternating section
followed by all documents in a single language,
and the size of the alternating section is gradually increased.
This is similar to \textit{shifting}, but with overlapping languages at the top instead of in the middle of the rank list. 
At the left of Figure~\ref{fig:diff-pattern}(d)
only the document at rank 1 is minority language, followed by minority language at ranks 1 and 3, and so on.
The right of the graph is identical to the right of Figure~\ref{fig:diff-pattern}(a).
These four patterns demonstrate that PEER scores match our intuition of fairness between languages. 
The next section evaluates real MLIR retrieval systems on two MLIR evaluation collections. 

\subsection{Assigning Languages to a Real Ranked List}

We used NeuCLIR'22 runs to create new synthetic runs
with relevant documents in the same positions,
but with languages assigned to the relevant documents either fairly or unfairly.
We randomly selected how many relevant documents would be assigned to each language,
and created that many language labels.
For each label we drew from a normal distribution
with either the same mean for the two languages (fair),
or different means (unfair).
We assigned a drawn number to each label,
sorted the labels by that number,
and assigned the labels to the relevant documents in the resulting order.
We did the same for the nonrelevant documents,
ensuring that each language was assigned at least 45\% of those documents.

Figures~\ref{fig:diff-pattern}(e) and (f) vary the sampling mean of the second language's relevant and nonrelevant documents, respectively,
while keeping a first language sampling mean of 1.0. 
The figures show that PEER captures fairness independently for each relevance level.
Since there are far fewer relevant documents,
the evidence for fairness is also weaker,
resulting in slower decay when changing the sampling mean for relevant documents.

\subsection{Real MLIR Systems}

We evaluate five MLIR systems, including query translation (QT) and document translation (DT) with BM25,
DT with English ColBERT~\cite{colbertv2},
and ColBERT-X models trained with English triples (ET) and multilingual translate-train (MTT)~\cite{lawrie2023neural},
on CLEF 2003 (German, Spanish, French, and English documents with English queries)
and NeuCLIR 2022 (Chinese, Persian, and Russian documents with English queries). 
For QT, English queries are translated into each document language
and monolingual search results from each language are fused by score. 
We report $\alpha$-nDCG, AWRF (with number of relevant documents as target distribution),
and the proposed PEER with rank cutoffs at 20 and 1000. 
Along with nDCG@20 and Recall@1000, we summarize the results in Table~\ref{tab:main}. 

Logically, merging ranked lists from each monolingual BM25 search with translated queries purely by scores
is inviting unfair treatment,
as scores from each language are incompatible with different query lengths and collection statistics~\cite{robertson2002trec}.
We observed this trend in both PEER and AWRF, %
while $\alpha$-nDCG strongly correlates with the effectiveness scores
and does not distinguish fairness. %

Neural MLIR models trained with only English text and transferred zero-shot to MLIR with European languages
exhibit a strong language bias compared to those trained with document languages \cite{lawrie2023neural, huang2023soft}.
PEER exhibits a similar trend in CLEF 2003, showing ColBERT-X ET is less fair than the MTT counterpart,
while AWRF is less sensitive. 
\citet{lawrie2023neural} show that preference for English documents in the ET model
causes this unfair treatment;
this suggests that MLIR tasks without the training language (English) in the document collection
would not suffer from such discrepancy. 
In fact, both PEER and AWRF indicate that MTT model is less fair among the three languages in NeuCLIR 2022, which is likely caused by the quality differences in machine translation~\cite{neuclir2022overview}. 

AWRF and PEER disagree on the comparison between English ColBERT on translated documents (DT) and ColBERT-X models. 
While AWRF suggests DT >> ColBERT
\footnote{The \textit{(ET)+ITD} setting in \citet{lawrie2023neural}.}
is fairer than ColBERT-X MTT in CLEF03, DT creates a larger difference among languages~\cite{lawrie2023neural}.
PEER, in contrast, aligns with prior analysis,
giving a lower score to DT >> ColBERT.
According to \citet{huang2023soft}, QT >> BM25 has a similar language bias compared to mDPR~\cite{tydiqa},
which was trained with English MS MARCO~\cite{msmarco}. 
PEER suggests a similar conclusion between QT >> BM25 and ColBERT-X ET,
which AWRF assigns a larger difference between the two with a rank cutoff of 20.

With a rank cutoff of 1000, AWRF strongly correlates with recall
(Pearson $r=0.93$ over both collections), while PEER does not (Pearson $r=-0.55$).
The 0.904 AWRF value (range 0-1) of DT >> ColBERT on CLEF03 suggests a fair system,
while the ranked list does not. 
This strong relationship shows that AWRF, with target distribution being the ratio of relevant documents,
is indeed measuring recall instead of fairness. 
While it is an artifact of the choice of target distribution,
the need to define a target distribution reduces the robustness of AWRF in measuring MLIR Fairness.

\section{Summary}

We propose measuring the Probability of Equal Expected Rank (PEER) for MLIR fairness. 
As PEER measures the weighted $p$-value of a non-parametric group hypothesis test,
it neither requires a target distribution nor makes distributional assumptions;
this makes the metric robust.
Through comparison to prior analytical work in MLIR Fairness,
we conclude that PEER captures the differences and nuances between systems better than other fairness metrics.

\bibliographystyle{ACM-Reference-Format}
\bibliography{sample-base}


\begin{thebibliography}{34}


\ifx \showCODEN    \undefined \def \showCODEN     #1{\unskip}     \fi
\ifx \showDOI      \undefined \def \showDOI       #1{#1}\fi
\ifx \showISBNx    \undefined \def \showISBNx     #1{\unskip}     \fi
\ifx \showISBNxiii \undefined \def \showISBNxiii  #1{\unskip}     \fi
\ifx \showISSN     \undefined \def \showISSN      #1{\unskip}     \fi
\ifx \showLCCN     \undefined \def \showLCCN      #1{\unskip}     \fi
\ifx \shownote     \undefined \def \shownote      #1{#1}          \fi
\ifx \showarticletitle \undefined \def \showarticletitle #1{#1}   \fi
\ifx \showURL      \undefined \def \showURL       {\relax}        \fi
\providecommand\bibfield[2]{#2}
\providecommand\bibinfo[2]{#2}
\providecommand\natexlab[1]{#1}
\providecommand\showeprint[2][]{arXiv:#2}

\bibitem[\protect\citeauthoryear{Arguello, Ferguson, Fine, Mitra, Zamani, and
  Diaz}{Arguello et~al\mbox{.}}{2021}]%
        {arguello2021tip}
\bibfield{author}{\bibinfo{person}{Jaime Arguello}, \bibinfo{person}{Adam
  Ferguson}, \bibinfo{person}{Emery Fine}, \bibinfo{person}{Bhaskar Mitra},
  \bibinfo{person}{Hamed Zamani}, {and} \bibinfo{person}{Fernando Diaz}.}
  \bibinfo{year}{2021}\natexlab{}.
\newblock \showarticletitle{Tip of the tongue known-item retrieval: A case
  study in movie identification}. In \bibinfo{booktitle}{\emph{Proceedings of
  the 2021 Conference on Human Information Interaction and Retrieval}}.
  \bibinfo{pages}{5--14}.
\newblock


\bibitem[\protect\citeauthoryear{Bajaj, Campos, Craswell, Deng, Gao, Liu,
  Majumder, McNamara, Mitra, Nguyen, et~al\mbox{.}}{Bajaj
  et~al\mbox{.}}{2016}]%
        {msmarco}
\bibfield{author}{\bibinfo{person}{Payal Bajaj}, \bibinfo{person}{Daniel
  Campos}, \bibinfo{person}{Nick Craswell}, \bibinfo{person}{Li Deng},
  \bibinfo{person}{Jianfeng Gao}, \bibinfo{person}{Xiaodong Liu},
  \bibinfo{person}{Rangan Majumder}, \bibinfo{person}{Andrew McNamara},
  \bibinfo{person}{Bhaskar Mitra}, \bibinfo{person}{Tri Nguyen},
  {et~al\mbox{.}}} \bibinfo{year}{2016}\natexlab{}.
\newblock \showarticletitle{Ms marco: A human generated machine reading
  comprehension dataset}.
\newblock \bibinfo{journal}{\emph{arXiv preprint arXiv:1611.09268}}
  (\bibinfo{year}{2016}).
\newblock


\bibitem[\protect\citeauthoryear{Biega, Gummadi, and Weikum}{Biega
  et~al\mbox{.}}{2018}]%
        {biega2018equity}
\bibfield{author}{\bibinfo{person}{Asia~J Biega}, \bibinfo{person}{Krishna~P
  Gummadi}, {and} \bibinfo{person}{Gerhard Weikum}.}
  \bibinfo{year}{2018}\natexlab{}.
\newblock \showarticletitle{Equity of attention: Amortizing individual fairness
  in rankings}. In \bibinfo{booktitle}{\emph{Proc. SIGIR}}.
  \bibinfo{pages}{405--414}.
\newblock


\bibitem[\protect\citeauthoryear{Braschler}{Braschler}{2001}]%
        {clef2001}
\bibfield{author}{\bibinfo{person}{Martin Braschler}.}
  \bibinfo{year}{2001}\natexlab{}.
\newblock \showarticletitle{{CLEF} 2001—Overview of Results}. In
  \bibinfo{booktitle}{\emph{Workshop of the Cross-Language Evaluation Forum for
  European Languages}}. Springer, \bibinfo{pages}{9--26}.
\newblock


\bibitem[\protect\citeauthoryear{Braschler}{Braschler}{2002}]%
        {clef2002}
\bibfield{author}{\bibinfo{person}{Martin Braschler}.}
  \bibinfo{year}{2002}\natexlab{}.
\newblock \showarticletitle{{CLEF} 2002—Overview of results}. In
  \bibinfo{booktitle}{\emph{Workshop of the Cross-Language Evaluation Forum for
  European Languages}}. Springer, \bibinfo{pages}{9--27}.
\newblock


\bibitem[\protect\citeauthoryear{Braschler}{Braschler}{2003}]%
        {clef2003}
\bibfield{author}{\bibinfo{person}{Martin Braschler}.}
  \bibinfo{year}{2003}\natexlab{}.
\newblock \showarticletitle{{CLEF} 2003--Overview of results}. In
  \bibinfo{booktitle}{\emph{Workshop of the Cross-Language Evaluation Forum for
  European Languages}}. Springer, \bibinfo{pages}{44--63}.
\newblock


\bibitem[\protect\citeauthoryear{Broder}{Broder}{2002}]%
        {broder2002taxonomy}
\bibfield{author}{\bibinfo{person}{Andrei Broder}.}
  \bibinfo{year}{2002}\natexlab{}.
\newblock \showarticletitle{A taxonomy of web search}. In
  \bibinfo{booktitle}{\emph{ACM Sigir forum}}, Vol.~\bibinfo{volume}{36}. ACM
  New York, NY, USA, \bibinfo{pages}{3--10}.
\newblock


\bibitem[\protect\citeauthoryear{Castillo}{Castillo}{2019}]%
        {castillo2019fairness}
\bibfield{author}{\bibinfo{person}{Carlos Castillo}.}
  \bibinfo{year}{2019}\natexlab{}.
\newblock \showarticletitle{Fairness and transparency in ranking}. In
  \bibinfo{booktitle}{\emph{ACM SIGIR Forum}}, Vol.~\bibinfo{volume}{52}. ACM
  New York, NY, USA, \bibinfo{pages}{64--71}.
\newblock


\bibitem[\protect\citeauthoryear{Chapelle, Metlzer, Zhang, and
  Grinspan}{Chapelle et~al\mbox{.}}{2009}]%
        {chapelle2009expected}
\bibfield{author}{\bibinfo{person}{Olivier Chapelle}, \bibinfo{person}{Donald
  Metlzer}, \bibinfo{person}{Ya Zhang}, {and} \bibinfo{person}{Pierre
  Grinspan}.} \bibinfo{year}{2009}\natexlab{}.
\newblock \showarticletitle{Expected reciprocal rank for graded relevance}. In
  \bibinfo{booktitle}{\emph{Proceedings of the 18th ACM conference on
  Information and knowledge management}}. \bibinfo{pages}{621--630}.
\newblock


\bibitem[\protect\citeauthoryear{Chen, Fisch, Weston, and Bordes}{Chen
  et~al\mbox{.}}{2017}]%
        {chen2017reading}
\bibfield{author}{\bibinfo{person}{Danqi Chen}, \bibinfo{person}{Adam Fisch},
  \bibinfo{person}{Jason Weston}, {and} \bibinfo{person}{Antoine Bordes}.}
  \bibinfo{year}{2017}\natexlab{}.
\newblock \showarticletitle{Reading Wikipedia to answer open-domain questions}.
  In \bibinfo{booktitle}{\emph{55th Annual Meeting of the Association for
  Computational Linguistics, ACL 2017}}. Association for Computational
  Linguistics (ACL), \bibinfo{pages}{1870--1879}.
\newblock


\bibitem[\protect\citeauthoryear{Choudhury and Deshpande}{Choudhury and
  Deshpande}{2021}]%
        {choudhury2021linguistically}
\bibfield{author}{\bibinfo{person}{Monojit Choudhury} {and}
  \bibinfo{person}{Amit Deshpande}.} \bibinfo{year}{2021}\natexlab{}.
\newblock \showarticletitle{How Linguistically Fair Are Multilingual
  Pre-Trained Language Models?}. In \bibinfo{booktitle}{\emph{Proceedings of
  the AAAI conference on artificial intelligence}}, Vol.~\bibinfo{volume}{35}.
  \bibinfo{pages}{12710--12718}.
\newblock


\bibitem[\protect\citeauthoryear{Clark, Choi, Collins, Garrette, Kwiatkowski,
  Nikolaev, and Palomaki}{Clark et~al\mbox{.}}{2020}]%
        {tydiqa}
\bibfield{author}{\bibinfo{person}{Jonathan~H. Clark}, \bibinfo{person}{Eunsol
  Choi}, \bibinfo{person}{Michael Collins}, \bibinfo{person}{Dan Garrette},
  \bibinfo{person}{Tom Kwiatkowski}, \bibinfo{person}{Vitaly Nikolaev}, {and}
  \bibinfo{person}{Jennimaria Palomaki}.} \bibinfo{year}{2020}\natexlab{}.
\newblock \showarticletitle{{TyDi QA}: A Benchmark for Information-Seeking
  Question Answering in Typologically Diverse Languages}.
\newblock \bibinfo{journal}{\emph{Transactions of the Association for
  Computational Linguistics}} (\bibinfo{year}{2020}).
\newblock


\bibitem[\protect\citeauthoryear{Clarke, Kolla, Cormack, Vechtomova, Ashkan,
  B{"u}ttcher, and MacKinnon}{Clarke et~al\mbox{.}}{2008}]%
        {Clarke2008NoveltyAD}
\bibfield{author}{\bibinfo{person}{Charles L.~A. Clarke},
  \bibinfo{person}{Maheedhar Kolla}, \bibinfo{person}{Gordon~V. Cormack},
  \bibinfo{person}{Olga Vechtomova}, \bibinfo{person}{Azin Ashkan},
  \bibinfo{person}{Stefan B{"u}ttcher}, {and} \bibinfo{person}{Ian MacKinnon}.}
  \bibinfo{year}{2008}\natexlab{}.
\newblock \showarticletitle{Novelty and diversity in information retrieval
  evaluation}. In \bibinfo{booktitle}{\emph{SIGIR}}.
\newblock


\bibitem[\protect\citeauthoryear{Diaz, Mitra, Ekstrand, Biega, and
  Carterette}{Diaz et~al\mbox{.}}{2020}]%
        {diaz2020evaluating}
\bibfield{author}{\bibinfo{person}{Fernando Diaz}, \bibinfo{person}{Bhaskar
  Mitra}, \bibinfo{person}{Michael~D Ekstrand}, \bibinfo{person}{Asia~J Biega},
  {and} \bibinfo{person}{Ben Carterette}.} \bibinfo{year}{2020}\natexlab{}.
\newblock \showarticletitle{Evaluating stochastic rankings with expected
  exposure}. In \bibinfo{booktitle}{\emph{Proceedings of the 29th ACM
  International Conference on Information \& Knowledge Management}}.
  \bibinfo{pages}{275--284}.
\newblock


\bibitem[\protect\citeauthoryear{Dwork, Hardt, Pitassi, Reingold, and
  Zemel}{Dwork et~al\mbox{.}}{2012}]%
        {dwork2012fairness}
\bibfield{author}{\bibinfo{person}{Cynthia Dwork}, \bibinfo{person}{Moritz
  Hardt}, \bibinfo{person}{Toniann Pitassi}, \bibinfo{person}{Omer Reingold},
  {and} \bibinfo{person}{Richard Zemel}.} \bibinfo{year}{2012}\natexlab{}.
\newblock \showarticletitle{Fairness through awareness}. In
  \bibinfo{booktitle}{\emph{Proceedings of the 3rd innovations in theoretical
  computer science conference}}. \bibinfo{pages}{214--226}.
\newblock


\bibitem[\protect\citeauthoryear{Huang, Zeng, Zamani, and Allan}{Huang
  et~al\mbox{.}}{2023}]%
        {huang2023soft}
\bibfield{author}{\bibinfo{person}{Zhiqi Huang}, \bibinfo{person}{Hansi Zeng},
  \bibinfo{person}{Hamed Zamani}, {and} \bibinfo{person}{James Allan}.}
  \bibinfo{year}{2023}\natexlab{}.
\newblock \showarticletitle{Soft Prompt Decoding for Multilingual Dense
  Retrieval}.
\newblock \bibinfo{journal}{\emph{arXiv preprint arXiv:2305.09025}}
  (\bibinfo{year}{2023}).
\newblock


\bibitem[\protect\citeauthoryear{Kassner, Dufter, and Sch{\"u}tze}{Kassner
  et~al\mbox{.}}{2021}]%
        {kassner2021multilingual}
\bibfield{author}{\bibinfo{person}{Nora Kassner}, \bibinfo{person}{Philipp
  Dufter}, {and} \bibinfo{person}{Hinrich Sch{\"u}tze}.}
  \bibinfo{year}{2021}\natexlab{}.
\newblock \showarticletitle{Multilingual LAMA: Investigating knowledge in
  multilingual pretrained language models}.
\newblock \bibinfo{journal}{\emph{arXiv preprint arXiv:2102.00894}}
  (\bibinfo{year}{2021}).
\newblock


\bibitem[\protect\citeauthoryear{Kruskal and Wallis}{Kruskal and
  Wallis}{1952}]%
        {kruskal1952use}
\bibfield{author}{\bibinfo{person}{William~H Kruskal} {and}
  \bibinfo{person}{W~Allen Wallis}.} \bibinfo{year}{1952}\natexlab{}.
\newblock \showarticletitle{Use of ranks in one-criterion variance analysis}.
\newblock \bibinfo{journal}{\emph{Journal of the American statistical
  Association}} \bibinfo{volume}{47}, \bibinfo{number}{260}
  (\bibinfo{year}{1952}), \bibinfo{pages}{583--621}.
\newblock


\bibitem[\protect\citeauthoryear{Kwon, Trivedi, Jansen, Surdeanu, and
  Balasubramanian}{Kwon et~al\mbox{.}}{2018}]%
        {kwon2018controlling}
\bibfield{author}{\bibinfo{person}{Heeyoung Kwon}, \bibinfo{person}{Harsh
  Trivedi}, \bibinfo{person}{Peter Jansen}, \bibinfo{person}{Mihai Surdeanu},
  {and} \bibinfo{person}{Niranjan Balasubramanian}.}
  \bibinfo{year}{2018}\natexlab{}.
\newblock \showarticletitle{Controlling information aggregation for complex
  question answering}. In \bibinfo{booktitle}{\emph{Advances in Information
  Retrieval: 40th European Conference on IR Research, ECIR 2018, Grenoble,
  France, March 26-29, 2018, Proceedings 40}}. Springer,
  \bibinfo{pages}{750--757}.
\newblock


\bibitem[\protect\citeauthoryear{Lawrie, MacAvaney, Mayfield, McNamee, Oard,
  Soldaini, and Yang}{Lawrie et~al\mbox{.}}{2023a}]%
        {neuclir2022overview}
\bibfield{author}{\bibinfo{person}{Dawn Lawrie}, \bibinfo{person}{Sean
  MacAvaney}, \bibinfo{person}{James Mayfield}, \bibinfo{person}{Paul McNamee},
  \bibinfo{person}{Douglas~W. Oard}, \bibinfo{person}{Luca Soldaini}, {and}
  \bibinfo{person}{Eugene Yang}.} \bibinfo{year}{2023}\natexlab{a}.
\newblock \showarticletitle{Overview of the TREC 2022 NeuCLIR Track}. In
  \bibinfo{booktitle}{\emph{Proceedings of the 31st Text REtrieval Conference}}
  (Gaithersburg, Maryland).
\newblock
\urldef\tempurl%
\url{https://arxiv.org/abs/2304.12367}
\showURL{%
\tempurl}


\bibitem[\protect\citeauthoryear{Lawrie, Yang, Oard, and Mayfield}{Lawrie
  et~al\mbox{.}}{2023b}]%
        {lawrie2023neural}
\bibfield{author}{\bibinfo{person}{Dawn Lawrie}, \bibinfo{person}{Eugene Yang},
  \bibinfo{person}{Douglas~W Oard}, {and} \bibinfo{person}{James Mayfield}.}
  \bibinfo{year}{2023}\natexlab{b}.
\newblock \showarticletitle{Neural Approaches to Multilingual Information
  Retrieval}. In \bibinfo{booktitle}{\emph{European Conference on Information
  Retrieval}}. Springer, \bibinfo{pages}{521--536}.
\newblock


\bibitem[\protect\citeauthoryear{Lee, Renear, and Smith}{Lee
  et~al\mbox{.}}{2006}]%
        {lee2006known}
\bibfield{author}{\bibinfo{person}{Jin~Ha Lee}, \bibinfo{person}{Allen Renear},
  {and} \bibinfo{person}{Linda~C Smith}.} \bibinfo{year}{2006}\natexlab{}.
\newblock \showarticletitle{Known-item search: Variations on a concept}.
\newblock \bibinfo{journal}{\emph{Proceedings of the american society for
  information science and technology}} \bibinfo{volume}{43},
  \bibinfo{number}{1} (\bibinfo{year}{2006}), \bibinfo{pages}{1--17}.
\newblock


\bibitem[\protect\citeauthoryear{Moffat and Zobel}{Moffat and Zobel}{2008}]%
        {moffat2008rank}
\bibfield{author}{\bibinfo{person}{Alistair Moffat} {and}
  \bibinfo{person}{Justin Zobel}.} \bibinfo{year}{2008}\natexlab{}.
\newblock \showarticletitle{Rank-biased precision for measurement of retrieval
  effectiveness}.
\newblock \bibinfo{journal}{\emph{ACM Transactions on Information Systems
  (TOIS)}} \bibinfo{volume}{27}, \bibinfo{number}{1} (\bibinfo{year}{2008}),
  \bibinfo{pages}{1--27}.
\newblock


\bibitem[\protect\citeauthoryear{Morik, Singh, Hong, and Joachims}{Morik
  et~al\mbox{.}}{2020}]%
        {morik2020controlling}
\bibfield{author}{\bibinfo{person}{Marco Morik}, \bibinfo{person}{Ashudeep
  Singh}, \bibinfo{person}{Jessica Hong}, {and} \bibinfo{person}{Thorsten
  Joachims}.} \bibinfo{year}{2020}\natexlab{}.
\newblock \showarticletitle{Controlling fairness and bias in dynamic
  learning-to-rank}. In \bibinfo{booktitle}{\emph{Proc. SIGIR}}.
  \bibinfo{pages}{429--438}.
\newblock


\bibitem[\protect\citeauthoryear{Rahimi, Shakery, and King}{Rahimi
  et~al\mbox{.}}{2015}]%
        {mulm}
\bibfield{author}{\bibinfo{person}{Razieh Rahimi}, \bibinfo{person}{Azadeh
  Shakery}, {and} \bibinfo{person}{Irwin King}.}
  \bibinfo{year}{2015}\natexlab{}.
\newblock \showarticletitle{Multilingual information retrieval in the language
  modeling framework}.
\newblock \bibinfo{journal}{\emph{Information Retrieval Journal}}
  \bibinfo{volume}{18}, \bibinfo{number}{3} (\bibinfo{year}{2015}),
  \bibinfo{pages}{246--281}.
\newblock


\bibitem[\protect\citeauthoryear{Robertson and Soboroff}{Robertson and
  Soboroff}{2002}]%
        {robertson2002trec}
\bibfield{author}{\bibinfo{person}{Stephen~E Robertson} {and}
  \bibinfo{person}{Ian Soboroff}.} \bibinfo{year}{2002}\natexlab{}.
\newblock \showarticletitle{The {TREC} 2002 {F}iltering Track Report.}. In
  \bibinfo{booktitle}{\emph{Proceedings of the Eleventh Text Retrieval
  Conference}}.
\newblock


\bibitem[\protect\citeauthoryear{Santhanam, Khattab, Saad-Falcon, Potts, and
  Zaharia}{Santhanam et~al\mbox{.}}{2022}]%
        {colbertv2}
\bibfield{author}{\bibinfo{person}{Keshav Santhanam}, \bibinfo{person}{Omar
  Khattab}, \bibinfo{person}{Jon Saad-Falcon}, \bibinfo{person}{Christopher
  Potts}, {and} \bibinfo{person}{Matei Zaharia}.}
  \bibinfo{year}{2022}\natexlab{}.
\newblock \showarticletitle{{C}ol{BERT}v2: Effective and Efficient Retrieval
  via Lightweight Late Interaction}. In \bibinfo{booktitle}{\emph{Proceedings
  of the 2022 Conference of the North American Chapter of the Association for
  Computational Linguistics: Human Language Technologies}}.
  \bibinfo{publisher}{Association for Computational Linguistics},
  \bibinfo{address}{Seattle, United States}, \bibinfo{pages}{3715--3734}.
\newblock


\bibitem[\protect\citeauthoryear{Sapiezynski, Zeng, E~Robertson, Mislove, and
  Wilson}{Sapiezynski et~al\mbox{.}}{2019}]%
        {sapiezynski2019quantifying}
\bibfield{author}{\bibinfo{person}{Piotr Sapiezynski}, \bibinfo{person}{Wesley
  Zeng}, \bibinfo{person}{Ronald E~Robertson}, \bibinfo{person}{Alan Mislove},
  {and} \bibinfo{person}{Christo Wilson}.} \bibinfo{year}{2019}\natexlab{}.
\newblock \showarticletitle{Quantifying the impact of user attention on fair
  group representation in ranked lists}. In \bibinfo{booktitle}{\emph{Companion
  Proceedings of WWW}}. \bibinfo{pages}{553--562}.
\newblock


\bibitem[\protect\citeauthoryear{Schulz-Hardt, Frey, L{\"u}thgens, and
  Moscovici}{Schulz-Hardt et~al\mbox{.}}{2000}]%
        {schulz2000biased}
\bibfield{author}{\bibinfo{person}{Stefan Schulz-Hardt},
  \bibinfo{person}{Dieter Frey}, \bibinfo{person}{Carsten L{\"u}thgens}, {and}
  \bibinfo{person}{Serge Moscovici}.} \bibinfo{year}{2000}\natexlab{}.
\newblock \showarticletitle{Biased information search in group decision
  making.}
\newblock \bibinfo{journal}{\emph{Journal of personality and social
  psychology}} \bibinfo{volume}{78}, \bibinfo{number}{4}
  (\bibinfo{year}{2000}), \bibinfo{pages}{655}.
\newblock


\bibitem[\protect\citeauthoryear{Singh and Joachims}{Singh and
  Joachims}{2018}]%
        {singh2018fairness}
\bibfield{author}{\bibinfo{person}{Ashudeep Singh} {and}
  \bibinfo{person}{Thorsten Joachims}.} \bibinfo{year}{2018}\natexlab{}.
\newblock \showarticletitle{Fairness of exposure in rankings}. In
  \bibinfo{booktitle}{\emph{Proc. KDD}}.
\newblock


\bibitem[\protect\citeauthoryear{White}{White}{2013}]%
        {white2013beliefs}
\bibfield{author}{\bibinfo{person}{Ryen White}.}
  \bibinfo{year}{2013}\natexlab{}.
\newblock \showarticletitle{Beliefs and biases in web search}. In
  \bibinfo{booktitle}{\emph{Proceedings of the 36th international ACM SIGIR
  conference on Research and development in information retrieval}}.
  \bibinfo{pages}{3--12}.
\newblock


\bibitem[\protect\citeauthoryear{Zehlike, Bonchi, Castillo, Hajian, Megahed,
  and Baeza-Yates}{Zehlike et~al\mbox{.}}{2017}]%
        {zehlike2017fa}
\bibfield{author}{\bibinfo{person}{Meike Zehlike}, \bibinfo{person}{Francesco
  Bonchi}, \bibinfo{person}{Carlos Castillo}, \bibinfo{person}{Sara Hajian},
  \bibinfo{person}{Mohamed Megahed}, {and} \bibinfo{person}{Ricardo
  Baeza-Yates}.} \bibinfo{year}{2017}\natexlab{}.
\newblock \showarticletitle{Fa* ir: A fair top-k ranking algorithm}. In
  \bibinfo{booktitle}{\emph{Proc. of CIKM}}.
\newblock


\bibitem[\protect\citeauthoryear{Zehlike, S{\"u}hr, Baeza-Yates, Bonchi,
  Castillo, and Hajian}{Zehlike et~al\mbox{.}}{2022}]%
        {zehlike2022fair}
\bibfield{author}{\bibinfo{person}{Meike Zehlike}, \bibinfo{person}{Tom
  S{\"u}hr}, \bibinfo{person}{Ricardo Baeza-Yates}, \bibinfo{person}{Francesco
  Bonchi}, \bibinfo{person}{Carlos Castillo}, {and} \bibinfo{person}{Sara
  Hajian}.} \bibinfo{year}{2022}\natexlab{}.
\newblock \showarticletitle{Fair Top-k Ranking with multiple protected groups}.
\newblock \bibinfo{journal}{\emph{Information processing \& management}}
  \bibinfo{volume}{59}, \bibinfo{number}{1} (\bibinfo{year}{2022}),
  \bibinfo{pages}{102707}.
\newblock


\bibitem[\protect\citeauthoryear{Zehlike, Yang, and Stoyanovich}{Zehlike
  et~al\mbox{.}}{2021}]%
        {zehlike2021fairness}
\bibfield{author}{\bibinfo{person}{Meike Zehlike}, \bibinfo{person}{Ke Yang},
  {and} \bibinfo{person}{Julia Stoyanovich}.} \bibinfo{year}{2021}\natexlab{}.
\newblock \showarticletitle{Fairness in ranking: A survey}.
\newblock \bibinfo{journal}{\emph{arXiv preprint arXiv:2103.14000}}
  (\bibinfo{year}{2021}).
\newblock


\end{thebibliography}

\end{document}